\documentclass[12pt,a4paper]{article}
\usepackage{amsmath,amsfonts,amssymb,bm}
\usepackage{graphicx} % do wczytywania rysunkow
\usepackage[hang,normalsize,bf]{subfigure} % rysuje kilka rysunkow
\begin{document}

\begin{center}
{\Huge \bf Resolved photon and multi-component model \\
for $\gamma^*$p and $\gamma^* \gamma^*$ scattering \\
at high energies \\}
\vspace {0.6cm}
{\large T. Pietrycki $^{1}$ and A. Szczurek $^{1,2}$\\}
\vspace {0.2cm}
$^{1}$ {\em Institute of Nuclear Physics\\
PL-31-342 Cracow, Poland\\}
$^{2}$ {\em University of Rzesz\'ow\\
PL-35-959 Rzesz\'ow, Poland\\}
\end{center}

\begin{abstract}
We generalize our previous model for $\gamma^* p$ scattering
to $\gamma \gamma$ scattering. In the latter case the number of
components naturally grows. When using the model parameters
from our previous $\gamma^* p$ analysis the model cross section for
$\gamma \gamma$ scattering is larger than the corresponding LEP2 
experimental data by more than a factor of two. However, performing 
a new simultaneous fit to $\gamma^* p$ and $\gamma \gamma$ total
cross section we can find an optimal set of parameters to describe 
both processes. We propose new measures of factorization breaking for $\gamma^* \gamma^*$ collisions and present results for our new model.
\end{abstract}

%\twocolumn
%---------------------
\section{Introduction}
%---------------------

In the last decade the photon-proton and photon-photon
reactions became a testing ground for different QCD-inspired
models. The dipole model was one of the most popular and successful
in this respect. In the simplest version of the model
only quark-antiquark Fock components of the photon are included
in order to describe the total cross sections.
In contrast, the more exclusive processes, like diffraction \cite{GBW_glue},
jet \cite{jets_resolved_photon} or heavy quark \cite{AS02} production, require
inclusion of higher Fock components of the photon. The higher Fock components 
can be of both perturbative and nonperturbative nature, and therefore
are rather difficult to include in a systematic manner.

In our recent publication \cite{PS03} we have constructed a simple hybrid model
which includes the resolved photon component in addition to
the quark-antiquark component. With a very small number of parameters
we were able to describe the HERA $\gamma^* p$ total cross section data 
with an accuracy similar to that of very popular dipole models
\cite{GBW,FKS99,KD00,MRS99,GLMN99}. The advantage of our model is that it
treats the total cross section and the exclusive processes
on the same footing.

The notion of the resolved photon is general and applies not
only to photon-proton collisions.
In the present paper we shall try to generalize our hybrid model
also to photon-photon collisions. Our approach is similar in the 
spirit to the approach of Ref.\cite{SS97} although the details 
differ considerably.
%---------------------------------
\section{Formulation of the model}
%---------------------------------
%-----------------------------------
\subsection{$\gamma^* p$ scattering}
%-----------------------------------

\begin{figure}[htb] % Figure 1 
\begin{center}
\includegraphics[width=10cm]{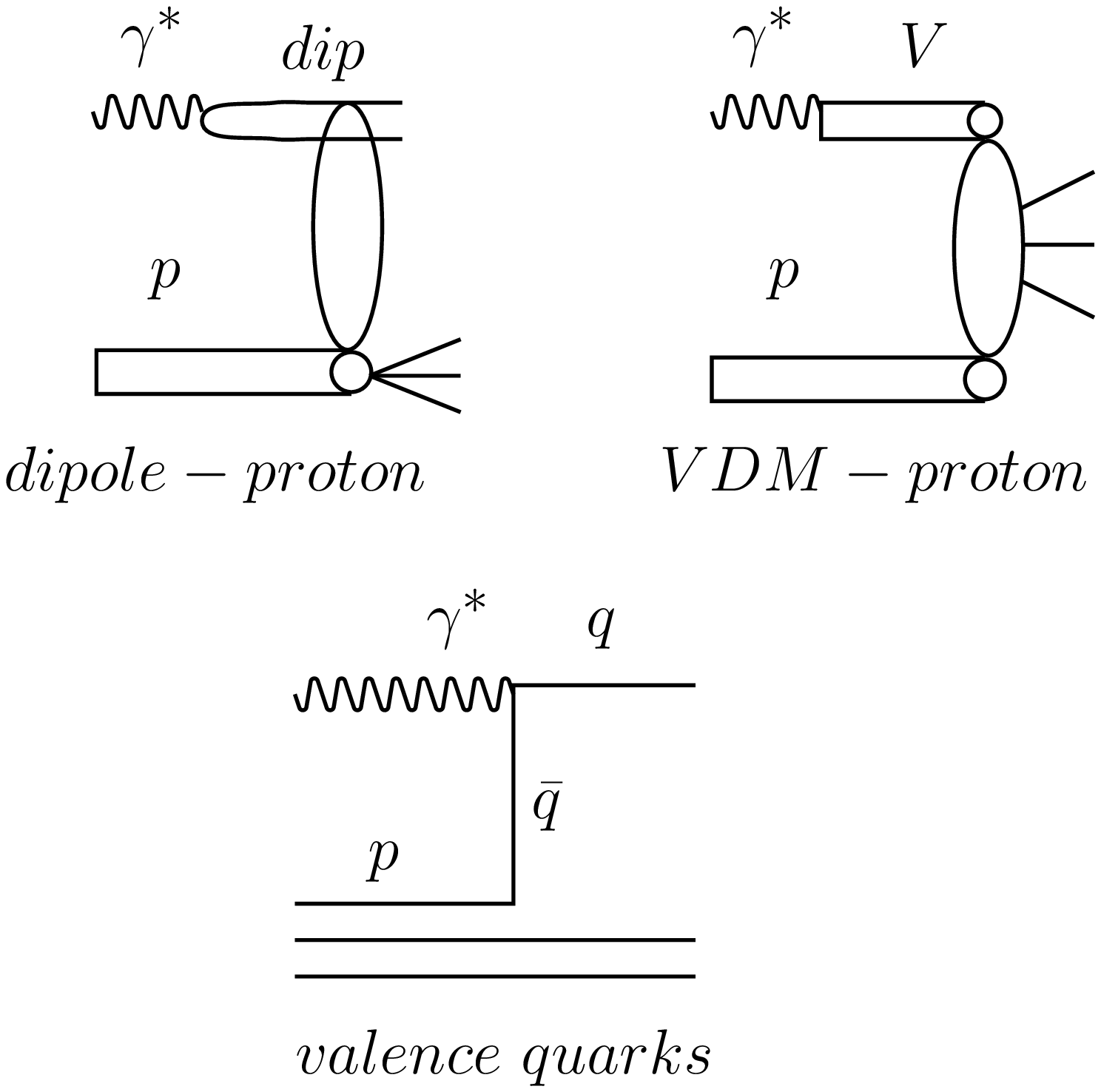}
\caption{\it
The graphical illustration of the multicomponent $\gamma^* $p 
scattering model.
\label{fig:fig2}
}
\end{center}
\end{figure}

First, let us recall our model for the total cross section 
for $\gamma^* p$ collisions. In this model the total cross section is a sum 
of three components,
\begin{eqnarray}
\sigma^{tot}_{\gamma^*N}(W,Q^2) &=&
   \sigma^{tot}_{dip}(W,Q^2) + 
\sigma^{tot}_{VDM}(W,Q^2) + \sigma^{tot}_{val}(W,Q^2)
\end{eqnarray}
where:
\begin{equation}
\sigma^{tot}_{dip}(W,Q^2) = \sum_q \int dz \int d^2\rho \;\sum_{T,L}
\left|\Psi^{T,L}_{\gamma^*\rightarrow q \bar q}(Q,z,\rho) \right|^2
\cdot \sigma_{(q \bar q)N}(x,\rho)  
\end{equation}
and
\begin{equation}
\sigma^{tot}_{VDM}(W,Q^2) = \sum_V \frac{4\pi}{\gamma^2_V}
\frac{M^4_V \sigma^{VN}_{tot}(W)}{(Q^2 + M_V^2)^2} \cdot (1 - x) . 
\end{equation}
All components of our model are illustrated graphically in Fig.\ref{fig:fig2}.
The last component becomes important only at large $x$, i.e. small W.

We take the simplest diagonal version of VDM with $\rho$, $\omega$ and
$\phi$ mesons included. As discussed recently in Ref.\cite{BS03} the
contributions of higher vector states are expected to be damped.
Above the meson-nucleon resonances it is reasonable to approximate
\begin{equation}
\sigma^{\rho N}_{tot}(W) = \sigma^{\omega N}_{tot}(W) =
\frac{1}{2}
 \left[ \sigma^{\pi^+ p}_{tot}(W) + \sigma^{\pi^- p}_{tot}(W) \right] \; ,
\label{sigma_VN}
\end{equation}
with a similar expression for $\sigma_{\phi p}^{tot}$ \cite{SU00}.
A simple Regge parametrization of the experimental pion-nucleon cross
section by Donnachie and Landshoff is used \cite{DL92}.
As in Ref. \cite{SU00} we take $\gamma$'s calculated from the leptonic decays
of vector mesons, including finite width corrections.
The factor (1-x) is meant to extend the VDM contribution towards
larger values of Bjorken $x$.

%------------------------------------------
\subsection{$\gamma^* \gamma^*$ scattering}
%------------------------------------------

\begin{figure}[htb] % Figure 2
\begin{center}
\includegraphics[width=10cm]{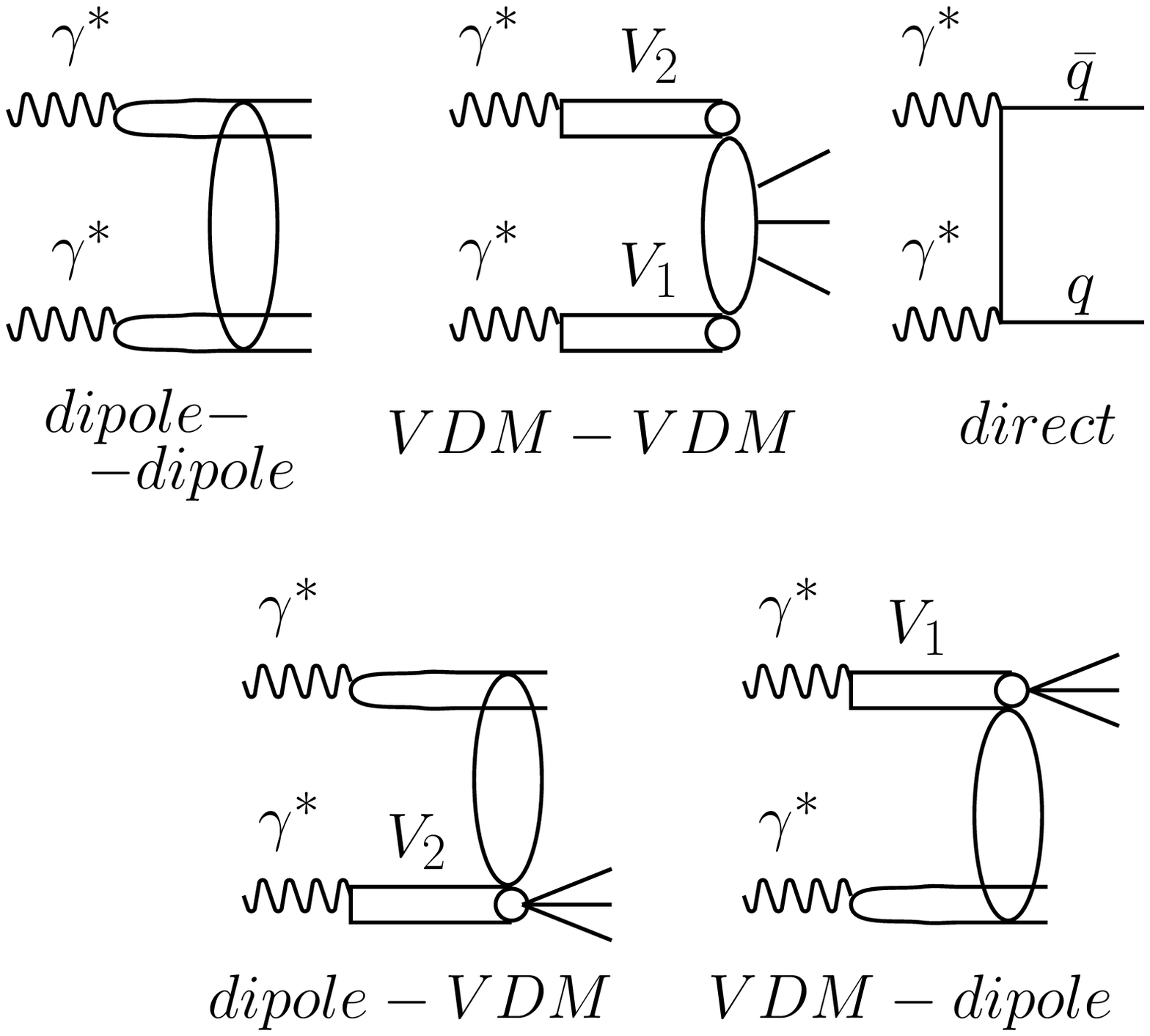}
\caption{\it
The graphical illustration of the multicomponent $\gamma^* \gamma^*$ 
scattering model.
\label{fig:fig1}
}
\end{center}
\end{figure}

In the same spirit, the total cross section for $\gamma^* \gamma^*$ scattering
can be written as a sum of the following five
terms (see Fig.\ref{fig:fig1}):
\begin{eqnarray}
\sigma^{tot}_{\gamma^*\gamma^*}(W,Q^2_1,Q^2_2)
&=&  \sigma^{tot}_{direct}(W,Q^2_1,Q^2_2) + \nonumber \\  
&+&  \sigma^{tot}_{dip-dip}(W,Q^2_1,Q^2_2) + \nonumber \\
&+&  \sigma^{tot}_{SR1}(W,Q^2_1,Q^2_2) + \nonumber \\
&+&  \sigma^{tot}_{SR2}(W,Q^2_1,Q^2_2) + \nonumber \\
&+&  \sigma^{tot}_{DR}(W,Q^2_1,Q^2_2).  \nonumber \\
\label{gg_expansion}
\end{eqnarray} 
The direct term, not possible in the case of photon-proton
scattering, is related to a new (as compared to the previous case) possibility
of $\gamma \gamma \to$ quark + antiquark process, and can be written
formally as a sum over quark flavours
\begin{equation}
\sigma^{tot}_{direct}(W,Q^2_1,Q^2_2) =  
\sum_f \sigma_{\gamma\gamma\rightarrow q_f \bar{q}_f}(W,Q^2_1,Q^2_2)
\; .  
\end{equation}
The corresponding formulae have been known for a long time 
and can be found in Ref.\cite{Budnev}.

If both photons fluctuate into perturbative quark-antiquark pairs,
the interaction is due to gluonic exchanges between quarks and antiquarks
represented in Fig.\ref{fig:fig1} by the blob.

Formally this component can be written in terms of the photon
perturbative "wave functions" and the cross section for the interaction
of both dipoles
\begin{eqnarray}
\sigma^{tot}_{dip-dip}(W,Q^2_1,Q^2_2)&=& 
\sum^{N_f}_{a,b=1}
\int_0^1dz_1\int d^2\rho_1|\Psi^a_T(z_1,\rho_1)|^2 \nonumber \\
&\cdot&\int_0^1dz_2\int d^2\rho_2
|\Psi^b_T(z_2,\rho_2)|^2 
\sigma^{a,b}_{dd}({\bar x}_{ab},\rho_1,\rho_2). \\ \nonumber   
\label{tot_dipdip}
\end{eqnarray}
The latter quantity is not well known. It can be easily calculated
in the simplest approach of two-gluon exchange. At high energies
such an approach cannot be sufficient, as gluonic ladders become
essential. Due to large degree of complexity a phenomenological
attitude seems indispensable.
In paper \cite{TKM02} a new phenomenological
parametrization for the azimuthal-angle averaged dipole-dipole
cross section has been proposed:
\begin{equation}
\sigma_{dd}^{a,b}(x_{ab},\rho_1,\rho_2) =
\sigma_0^{a,b}
\left [
 1 - \exp\left(- \frac{\rho_{eff}^2}{4 R_0^2(x_{ab})}  \right)
\right ] 
\cdot S_{thresh}(x_{ab}) \; .
\label{saturation_parametrization}
\end{equation}
Here
\begin{equation}
x_{ab} = \frac{\frac{m_a^2}{z_1}+\frac{m_a^2}{1-z_1}
+\frac{m_b^2}{z_2}+\frac{m_b^2}{1-z_2}+Q_1^2+Q_2^2}
{W^2+Q_1^2+Q_2^2} \; 
\label{x_ab}
\end{equation}
and
\begin{equation}
R_0(x_{ab}) = \frac{1}{Q_0} 
\left(
\frac{x_{ab}}{x_0}
\right)
^{-\lambda/2}  \; .
\end{equation}
Our formula for $x_{ab}$ is different from the one used in 
Ref.\cite{TKM02}. As discussed in Ref.\cite{AS02} our formula 
provides correct behaviour at threshold energies.

In order to take into account threshold effects for the production
of $q \bar q q' \bar q'$ an extra phenomenological function
has been introduced \cite{TKM02}
\begin{equation}
S_{thresh}(x_{ab}) = (1 - x_{ab})^5 
\label{threshold_function}
\end{equation}
which is set to zero if $x_{ab} > 1$.
Different prescriptions for $\rho_{eff}$ have been considered
in Ref.\cite{TKM02}, with $\rho_{eff}^2 = \frac{\rho_1^2 \rho_2^2}{\rho_1^2 +
  \rho_2^2}$ being probably the best choice \cite{TKM02}.
Following our philosophy of explicitly including the
nonperturbative resolved photon, in photon-photon collisions completely 
new terms must be included (the last two diagrams in
Fig.\ref{fig:fig1}).
If one of the photons fluctuates into a quark-antiquark dipole and
the second photon fluctuates into a vector meson, or vice versa,
we shall call such components single resolved components.
In $\gamma \gamma$ scattering there are two such components: 
\begin{eqnarray}
\sigma^{tot}_{SR1}(W,Q^2_1,Q^2_2) &=& 
\int d^2\rho_2\int dz_2 \sum_{V_1}\frac{4\pi}{f^2_{V_1}}
   \left(\frac{m^2_{V_1}}{m^2_{V_1}+Q^2_1}\right)^2 \cdot \nonumber \\ 
&\cdot& \left|\Psi(\rho_2,z_2,Q^2_2)\right|^2 
\sigma^{tot}_{V_1d}(W,Q^2_2)  \; , 
\end{eqnarray} 
\begin{eqnarray}
\sigma^{tot}_{SR2}(W,Q^2_1,Q^2_2) &=& 
\int d^2\rho_1\int dz_1 \sum_{V_2}\frac{4\pi}{f^2_{V_2}}
   \left(\frac{m^2_{V_2}}{m^2_{V_2}+Q^2_2}\right)^2 \cdot \nonumber \\ 
&\cdot& \left|\Psi(\rho_1,z_1,Q^2_1)\right|^2 
\sigma^{tot}_{V_2d}(W,Q^2_1) \; .  
\end{eqnarray} 
In the formulae above:
\begin{equation}
\sigma_{V_i d}^{tot}(W,Q^2) = \sigma_0 \left(1-\mathrm{exp}
\left(-\frac{\rho_i^2}{4R_0^2(x_g)}\right) \right)\cdot S_{thresh}
\end{equation}
where
\begin{equation}
R_0(x_g) = \frac{1}{Q_0}\cdot \left(\frac{x_g}{x_0}\right)^{\lambda/2}
\end{equation}
and, to a good approximation,
\begin{equation}
x_g = \frac{M_{qq}^2+Q^2}{W^2+Q^2}
\end{equation}
with
\begin{equation}
M_{qq}=\frac{m_f^2}{z(1-z)} \; ,
\end{equation}
where $m_f$ is quark effective mass. In the present calculation
we take $m_f = m_0$
for $u / {\bar u}$ and $d / {\bar d}$ (anti)quarks and $m_f = m_0 +$ 0.15
GeV for $s/{\bar s}$ (anti)quarks.

If each of the photons fluctuates into a vector meson the
corresponding component will be called double resolved. 
\footnote{In some early works in the literature this was 
considered as the only component to the photon-photon 
total cross section (see for instance \cite{GS82}).} 
The corresponding cross section reads formally:
\begin{eqnarray}
\sigma^{tot}_{DR}(W,Q^2_1,Q^2_2) &=&
\sum_{V_1 V_2} \frac{4\pi}{f^2_{V_1}}
  \left(\frac{m^2_{V_1}}{m^2_{V_1}+Q^2_1}\right)^2 \cdot \nonumber \\
&\cdot& \frac{4\pi}{f^2_{V_2}}
  \left(\frac{m^2_{V_2}}{m^2_{V_2}+Q^2_2}\right)^2
  \sigma^{tot}_{V_1 V_2}(W) \; .  
\end{eqnarray} 
%
%% Here the off-shell form factor $F_{off}(Q^2,m^2_V)$ describes the coupling
%% of the virtual photon to virtual vector meson.
%% This form factor guarantees the required condition
%% $F_{off}(-m^2_V,m^2_V) = 1 $.

The total cross section for $V_1$-$V_2$ scattering
must be modeled. In the following we shall assume
Regge factorization and use a simple parametrization which
fits the world experimental data for hadron-hadron
total cross sections \cite{DL92}. It was demonstrated
recently that in the case of the total cross sections
the absorption corrections violate the factorization
only mariginally \cite{SNS01}.
Assuming factorization and neglecting the off-diagonal terms
due to the $a_2$-reggeon echange we obtain a simple
and economical form
\begin{equation}
\sigma_{V_1 V_2}^{tot}(W) =
A_{R}  \left( \frac{s}{s_0} \right)^{\alpha_{R}-1} +
A_{IP} \left( \frac{s}{s_0} \right)^{\alpha_{IP}-1}
\label{V1_V2_total_cross_section}
\end{equation}
with $A_{R} = 13.2\; mb$ and $A_{IP} = 8.56\; mb$,
$\alpha_{R}=0.5$, $\alpha_{IP}=1.08$, $s = W^2 $, $s_0 = 1\;GeV^2$ .

%----------------
\section{Results}
%----------------

\begin{figure}[htb] % Figure 3
\begin{center}
\includegraphics[width=9cm]{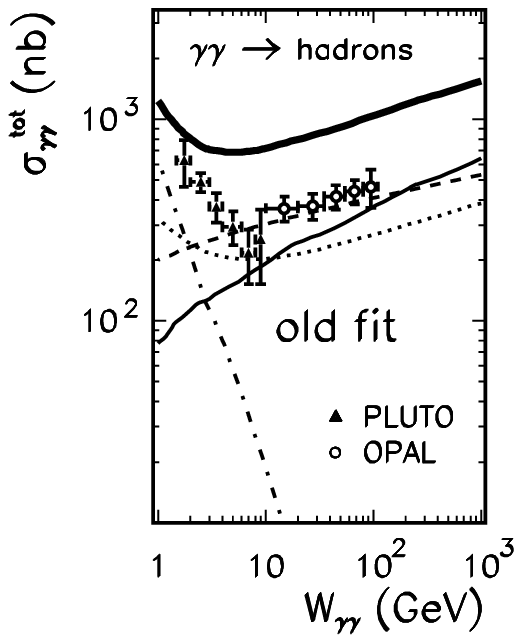}
\caption{\it  
The total $\gamma \gamma$ cross section as a function 
of photon-photon energy with parameters from Ref.\cite{PS03}.
The experimental data are from \cite{PLUTO,OPAL}.
\label{fig:gg_tot}
}
\end{center}
\end{figure}

In Ref.\cite{PS03} we have adjusted the parameters of our model
to $\gamma^* p$ collisions. Let us try to use these parameters
to describe $\gamma \gamma$ total cross section.
In Fig.\ref{fig:gg_tot} we present the total cross section
as a function of center-of-mass energy.
The sum of all components of Fig.\ref{fig:fig1} (thick-solid line) exceeds
the experimental data by a factor of two or even more. The individual
components are shown explicitly as well. The direct component
(dash-dotted line) dominates at low energies only. At high energies 
the dipole-dipole (thin-solid line), single-resolved (dashed line) and 
double-resolved (dotted line) components are of comparable size.
The overestimation of the experimental data suggests a double-counting.

Let us try to recapitulate the assumptions and/or approximations
used in obtaining the formulae of the previous section.
First of all it was assumed that the coupling constants
responsible for the transition of photons into vector mesons
are the same as those obtained from the leptonic decays of vector
mesons, i.e. the on-shell approximation was used.
In our case we need the corresponding coupling constants rather at
$Q^2$ = 0 and not on the meson mass shell ($Q^2 = m_V^2$).
In principle, there can be a weak modification by a $Q^2$-dependent
function. We replace $\frac{4\pi}{f^2_{V_i}}\rightarrow 
\frac{4\pi}{f^2_{V_i}}F_{off}(Q^2,m^2_{V_i}) $ and propose 
to parametrize the effect of extrapolation
from meson mass shell to $Q^2$ = 0 by means of the following
form factor:
%------------------
\begin{equation}
F_{off}(Q^2,m^2_{V_i}) = \mathrm{exp}\left(-\frac{(Q^2 + m^2_{V_i})}
{2\Lambda_E^2} \right)
\; . 
\label{F_off}
\end{equation}
%--------------------
The parameter $\Lambda_E$ is a new nonperturbative parameter 
of our new model.
Secondly, the ``photon-wave functions'' commonly used in the 
literature allow for large quark-antiquark dipoles. 
This is a nonperturbative region
where the pQCD is not expected to work. Furthermore this is
a region which is probably taken into account in the resolved
photon components as explicit vector mesons.
Therefore large-size dipoles must be removed from
the photon wave functions. We propose the following
modification of the ``perturbative'' photon wave function:
%------------------------------------
\begin{equation}
\left|\Psi(\rho,z,Q^2) \right|^2 \rightarrow \left|\Psi(\rho,z,Q^2) \right|^2
\mathrm{exp}\left(- \frac{\rho}{\rho_0} \right) \: ,
\label{dipole_size_modification}
\end{equation}
%------------------------------------
which effectively suppresses large quark-antiquark dipoles.

In the following we shall try to find the parameters
$\Lambda_E$ and $\rho_0$ by fitting our modified model formula to 
the experimental data.
The $\gamma \gamma$ data is not sufficient for this purpose
as different combinations of the two parameters lead
to equally good description. Therefore we are forced to perform a
new fit of the model parameters to both $\gamma^* p$
and $\gamma \gamma$ scattering.

Naively one could try to adjust the new parameters in Eq.(\ref{F_off})
and Eq.(\ref{dipole_size_modification}) to describe the photon-photon
data only. However, internal consistency would require associated modifications
in $\gamma^* p$ collisions. It is obvious that such modifications would
destroy the nice agreement with the HERA data \cite{HERA_data} as 
obtained in Ref.\cite{PS03}.
It becomes clear that a new simultaneous fit of the extended model
to both $\gamma^* p$ and $\gamma \gamma$ is unavoidable.
It is not clear {\it a priori} that a good quality fit is possible at all.

In order to quantify the quality of the simultaneous fit we propose
the following simple measure of fit quality: 
%----------------------------------------------------------
\begin{equation}
\chi^2_{eff} = \frac{\frac{\chi^{2}_{\gamma^* p}}
{N_{\gamma^* p}} + 
\frac{\chi^{2}_{\gamma \gamma}}{N_{\gamma \gamma}}}
{2} \; .
\label{chi2}
\end{equation}
%-----------------------------------------------------

This is a bit {\it ad hoc} statistically, but treats the $\gamma^* p$
and $\gamma \gamma$ processes with the same weight which seems reasonable
in view of large disproportions of the $\gamma^* p$ and $\gamma \gamma$
data sets.
In the present fit in addition to the HERA \cite{HERA_data} data
for $\gamma^*$p scattering we include also the PLUTO \cite{PLUTO} and OPAL
\cite{OPAL} collaboration data for $\gamma \gamma$ scattering.

In Tables 1, 2 we have collected the values of minimal
standard $\chi^2$ for different pairs of the newly
defined parameters of the extended model: $\rho_0$ and $\Lambda_E$.
Each value of $\chi^2$ is supplemented with the values of
the remaining model parameters ($\sigma_0$, $x_0$ and $\lambda$)
which we have not presented in the table for clarity. A rather good
description of both processes can be obtained.
However, the smallest values of $\chi^2$ for both processes are situated
in different parts of the two tables.
In Table 3 we display the effective $\chi^2$ defined by Eq.(\ref{chi2}).
Here the minimal value of the proposed measure $\chi^2_{eff}$ is at 
$\rho_0$ = 5.0 GeV$^{-1}$ and $\Lambda_E$ = 1 GeV for which $\chi_{eff}^2$ = 1.7.

%----------------------------------------------------
\begin{table}
\caption{$\chi^2$ in $\gamma^*p$ scattering}
\begin{center}
\begin{tabular}{|cl|rrrr|}
\hline
        &        &&\;\;$\Lambda_E$ &&\\  
        &        & 0.5 & 1.0 & 2.0 &$\infty$\\
\hline               
        &  1.0   &104.0&59.0&23.0&11.0\\
        &  2.0   & 51.0&22.0& 4.7& 2.4\\
        &  3.0   & 28.0& 8.7& 2.4& 2.5\\      
$\rho_0$&  4.0   & 19.0& 5.0& 2.3& 3.0\\
        &  5.0   & 10.0& 2.1& 2.4& 3.0\\
        &  6.0   &  7.2& 1.8& 2.5& 3.3\\
        &$\infty$&  2.1& 2.2& 4.6& 8.3\\
\hline  
\end{tabular}
\end{center}
\end{table}
%--------------------------------------------------
\begin{table}
\caption{ $\chi^2$ in $\gamma \gamma$ scattering}
\begin{center}
\begin{tabular}{|cl|rrrr|}
\hline
        &         &&\;\;$\Lambda_E$ &&\\
        &         &  0.5 & 1.0 & 2.0 & $\infty$ \\
\hline
        &  1.0    &14.0 & 5.2 & 0.7 & 1.2  \\
        &  2.0    &12.0 & 2.3 & 1.8 & 5.0  \\
        &  3.0    & 9.8 & 1.1 & 4.4 & 8.1  \\      
$\rho_0$&  4.0    & 7.9 & 1.0 & 2.6 &12.0  \\
        &  5.0    & 6.8 & 1.6 & 8.9 &13.0  \\
        &  6.0    & 5.7 & 2.4 &11.0 &16.0  \\
        &$\infty$ & 1.5 &19.0 &43.0 &59.0  \\
\hline  
\end{tabular}
\end{center}
\end{table}
%---------------------------------------------------------
\begin{table}
\caption{ $\chi^2_{eff}$ in $\gamma^*$p and $\gamma \gamma$ scattering}
\begin{center}
\begin{tabular}{|cl|rrrr|}
\hline
        &         &&\;\;$\Lambda_E$ &&\\
        &         & 0.5 & 1.0 & 2.0 & $\infty$ \\
\hline
        &  1.0    &59.0&32.0&12.0& 6.1 \\
        &  2.0    &32.0&12.0& 3.3& 3.7 \\
        &  3.0    &19.0& 4.9& 3.4& 5.3 \\      
$\rho_0$&  4.0    &14.0& 3.0& 2.5& 7.5 \\
        &  5.0    & 8.4& 1.3& 5.7& 8.0 \\
        &  6.0    & 6.5& 2.1& 6.5& 9.7 \\
        &$\infty$ & 1.8&11.0&24.0&34.0 \\
\hline  
\end{tabular}
\end{center}
\end{table}                                                                    
%-------------------------------------------

\begin{figure}[htb] % Figure 4  
\begin{center}
\includegraphics[width=9cm]{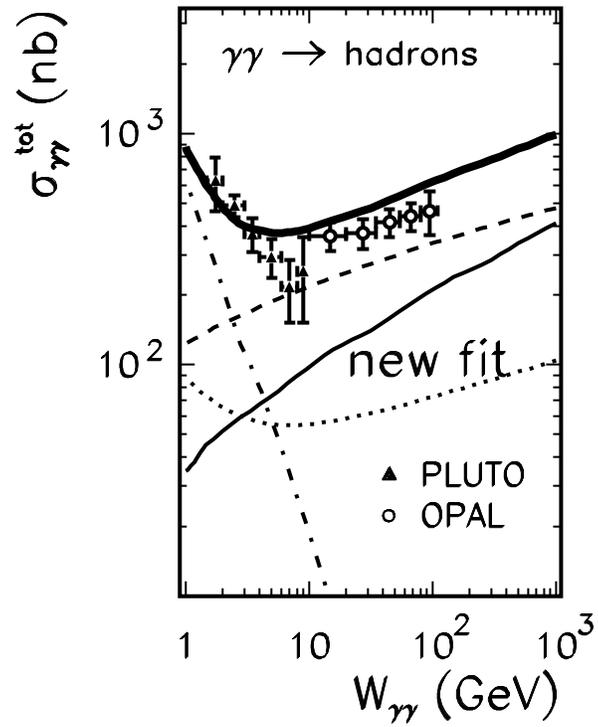}
\caption{\it
The total $\gamma \gamma$ cross section as a function 
of photon-photon energy with the new set of parameters.
The experimental data are from \cite{PLUTO,OPAL}.
\label{fig:gg_tot18}
}
\end{center}
\end{figure}

\begin{figure}[htb] % Figure 5
\begin{center}
\includegraphics[width=10cm]{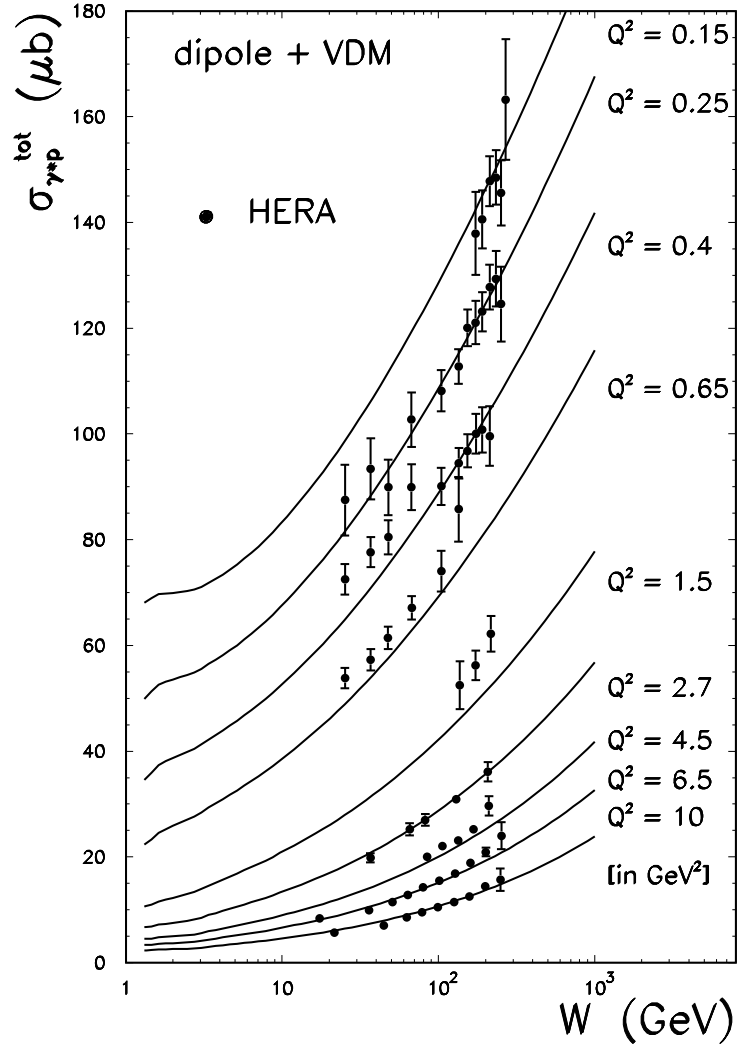}
\caption{\it
The total $\gamma^* p$ cross section as a function 
of photon-proton energy. The experimental HERA data 
are from \cite{HERA_data}.
\label{fig:hera}
}
\end{center}
\end{figure}

In Fig.\ref{fig:gg_tot18} we show the resulting total
cross section for the photon-photon scattering together with the
experimental data of the PLUTO (solid triangles) and OPAL (open circles)
collaborations. We also show the individual contributions of different
processes from Fig.\ref{fig:fig1}. Please note that the relative size of the contributions
has changed when compared to Fig.\ref{fig:gg_tot}. Now the sum of the 
so-called single resolved components
dominates in the broad range of center-of-mass energies. It is worth
stressing in this context that these components are included here
for the first time. When compared to Fig.\ref{fig:gg_tot} the
double resolved component is now much weaker and constitutes 10-15 \%
of the total cross section only.
For completeness in Fig.\ref{fig:hera} we show the analogous description
of the $\gamma^* p$ data. The agreement with the HERA data is similar 
as in our previous paper \cite{PS03}.

%-------------------------------
\section{Factorization breaking}
%-------------------------------

In data processing, in particular in extrapolations to small 
photon virtualities one often assumes the following relation
%-------------------------------------------------
\begin{equation}
\sigma_{\gamma^* \gamma^*}^{tot}(W,Q_1^2,Q_2^2)
= \Omega(Q_1^2) \cdot \Omega(Q_2^2) \cdot \sigma(W)
\label{factorization}
\end{equation}
%--------------------------------------------------
known as factorization. This relation is strictly true
for single-pole double-resolved VDM components
and means total decorrelation of $Q_1^2$ and $Q_2^2$.
In the following we shall consider two quantities which
measure factorization breaking.

The first one reads
%--------------------------------------------------
\begin{equation}
f_{fb}^{(1)}(W,Q_1^2,Q_2^2) \equiv
\frac{\sigma_{\gamma^* \gamma^*}(W,Q_1^2,0) \;
      \sigma_{\gamma^* \gamma^*}(W,0,Q_2^2)}
     {\sigma_{\gamma^* \gamma^*}(W,Q_1^2,Q_2^2) \;
      \sigma_{\gamma^* \gamma^*}(W,0,0)} \; .
\label{f_fb1}
\end{equation}
%-------------------------------------------------
For the factorized Ansatz (\ref{factorization}) $f_{fb}^{(1)}$ = 1.
This quantity may be difficult to measure at present as it requires knowledge
of the cross section for real photons, which is not possible
with present $e^+ e^-$ colliders and the detectors used.
We hope this quantity can be used in the future with the help 
of the photon-photon option at TESLA \cite{TESLA_design_report}.

The second quantity \footnote{A similar quantity has been
used to study factorization breaking of a color dipole BFKL approach
\cite{NSZ02} 
to highly virtual photon - highly virtual photon
scattering} is
%--------------------------------------------------
\begin{equation}
f_{fb}^{(2)}(W,Q_1^2,Q_2^2) \equiv
\frac{\sigma_{\gamma^* \gamma^*}(W,Q_1^2,Q_1^2) \;
      \sigma_{\gamma^* \gamma^*}(W,Q_2^2,Q_2^2)}
     {\sigma_{\gamma^* \gamma^*}(W,Q_1^2,Q_2^2) \;
      \sigma_{\gamma^* \gamma^*}(W,Q_2^2,Q_1^2)}  \; .
\label{f_fb2}
\end{equation}
%---------------------------------------------------
As in the previous case it is easy to check that with the factorized
Ansatz (\ref{factorization}) $f_{fb}^{(2)}$ = 1.
The effect of factorization breaking is limited
through the following normalization condition
%---------------------------------
\begin{equation}
f_{fb}^{(2)}(W,Q^2,Q^2) = 1  \; .
\label{f_fb2_norma}
\end{equation}
%--------------------------------
Therefore it becomes clear that this quantity becomes interesting
if $Q_1^2 \gg Q_2^2$ or $Q_1^2 \ll Q_2^2$.
The second quantity mesures formally (de)correlations of both photons
virtualities. In principle, this quantity can be used in
the analysis of existing experimental data from DESY, SLAC or LEP.

\begin{figure}[htb] % Figure 6
%\begin{center}
\subfigure[]{\label{sig_w10}
    \includegraphics[width=7cm]{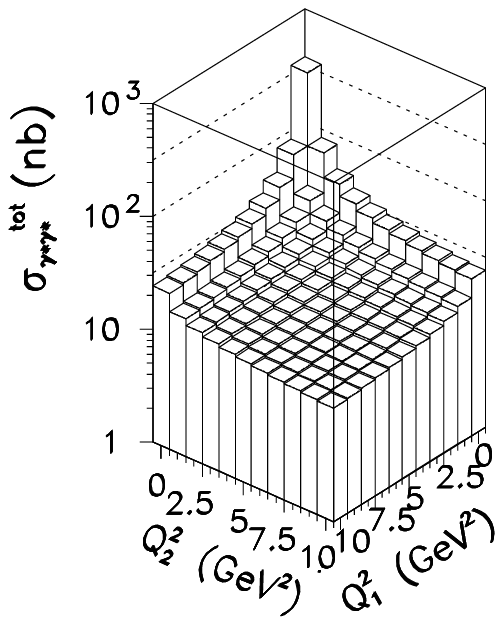}}
\subfigure[]{\label{sig_w100}
    \includegraphics[width=7cm]{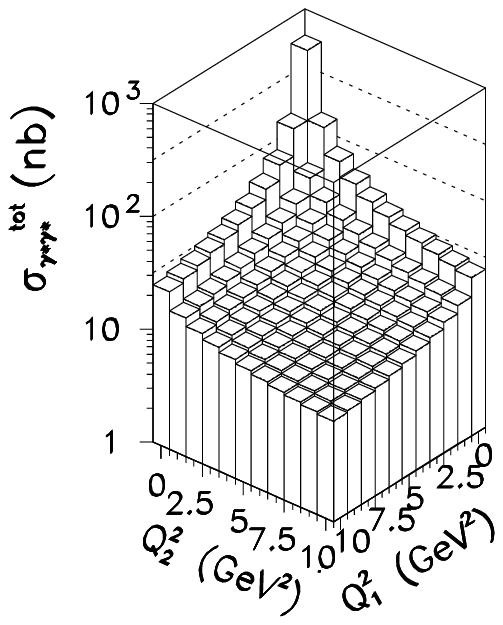}}
\caption{\it
The maps of the total $\gamma^* \gamma^*$ cross section 
as a function of both photon virtualities $Q_1^2$ and $Q_2^2$ 
for $W = 10$ GeV (left panel) and $W = 100$ GeV (right panel).
\label{fig:cs_tot_w}
}
%\end{center}
\end{figure}

Both quantities proposed for measuring factorization breaking require knowledge
of the total cross section not only for the real photons but also for the virtual 
ones. Before we present the quantities in question we wish to display the
total photon-photon cross section as a function of both photon virtualities.
In Fig.\ref{fig:cs_tot_w} we show the corresponding maps for two quite 
different energies $W =$ 10 GeV and $W =$ 100 GeV in measurable range of photon
virtualities. Two observations can be made here. First, the two maps look
rather similar. Secondly, fast fall-off is observed at photon
virtualities 0 $< Q^2 <$ 1 GeV$^2$, with further decrease being much softer.

\begin{figure}[htb] % Figure 7
%\begin{center}
 \subfigure[]{\label{fb1_w10}
    \includegraphics[width=7cm]{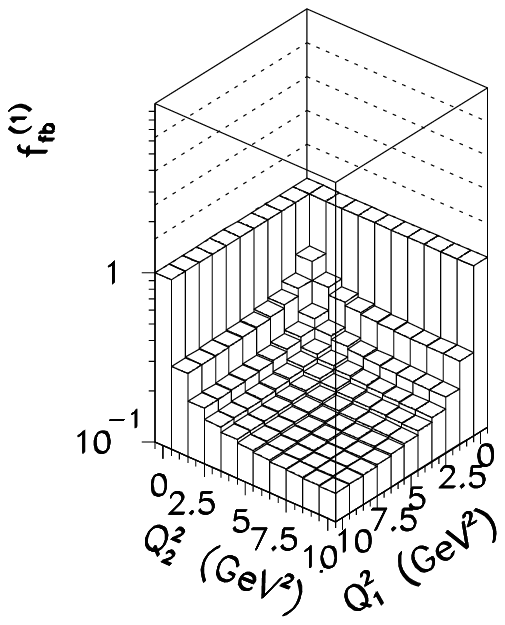}}
 \subfigure[]{\label{fb1_w100}
    \includegraphics[width=7cm]{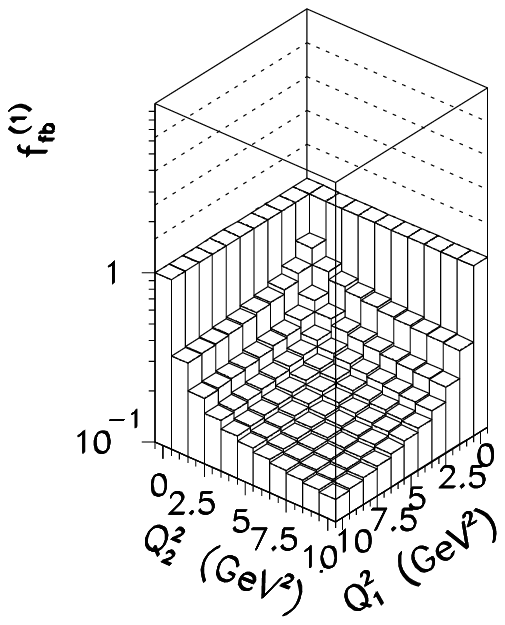}}
\caption{\it
The maps of the factorization-breaking function $f_{fb}^{(1)}$ 
as a function of both photon virtualities $Q_1^2$ and $Q_2^2$ for
$W = 10$ GeV (left panel) and $W = 100$ GeV (right panel).
\label{fig:fb1_w}
}
%\end{center}
\end{figure}

\begin{figure}[htb] % Figure 8
%\begin{center}
 \subfigure[]{\label{fb1_w10}
    \includegraphics[width=7cm]{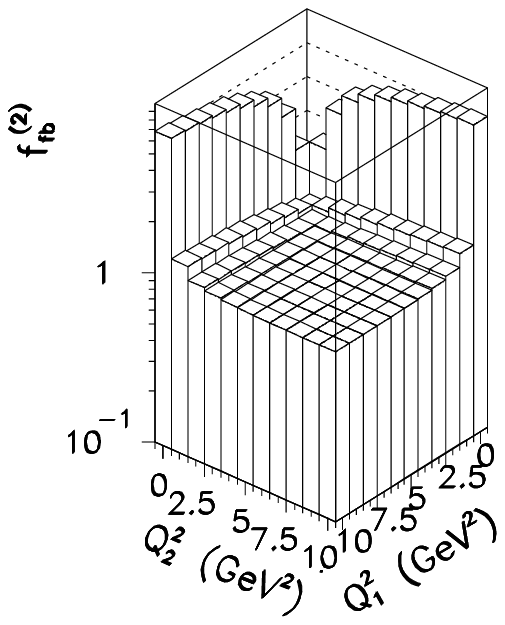}}
 \subfigure[]{\label{fb1_w100}
    \includegraphics[width=7cm]{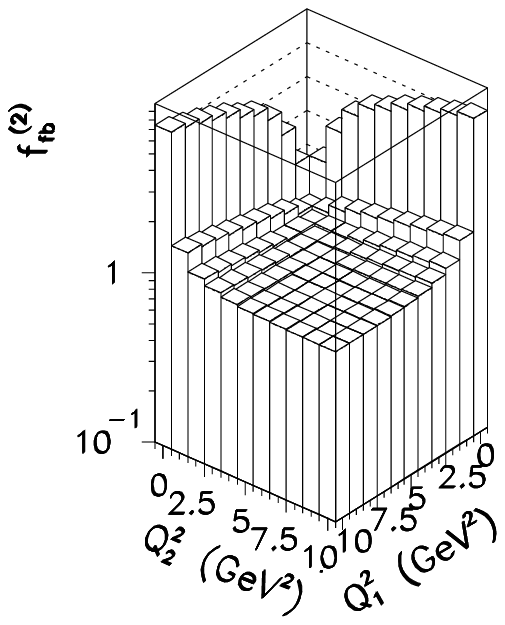}}
\caption{\it
The maps of the factorization-breaking function $f_{fb}^{(2)}$ 
as a function of both photon virtualities $Q_1^2$ and $Q_2^2$ for
$W = 10$ GeV (left panel) and $W = 100$ GeV (right panel).
\label{fig:fb2_w}
}
%\end{center}
\end{figure}

The factorization-breaking function $f_{fb}^{(1)}$ is shown in Fig.\ref{fig:fb1_w}
as a function of both photon virtualities $Q_1^2$ and $Q_2^2$ for W = 10 GeV 
(left panel) and W = 100 GeV (right panel).
According to the definition (\ref{f_fb1}) at $Q_1^2 = 0$ or $Q_2^2 = 0$ we have
$f_{fb}^{(1)}$ = 1. The rapid variation of the function is not best
represented by our rough grid.
For completeness the second proposed function is shown
in Fig.\ref{fig:fb2_w}. By definition this time (see Eq.(\ref{f_fb2}))
we have $f_{fb}^{(2)}$ = 1 when $Q_1^2 = Q_2^2$. As in the previous case
fast variation occurs at small photon virtualities. 

\begin{figure}[htb] % Figure 9
 \subfigure[]{\label{f1_q2_w10}
    \includegraphics[width=7cm]{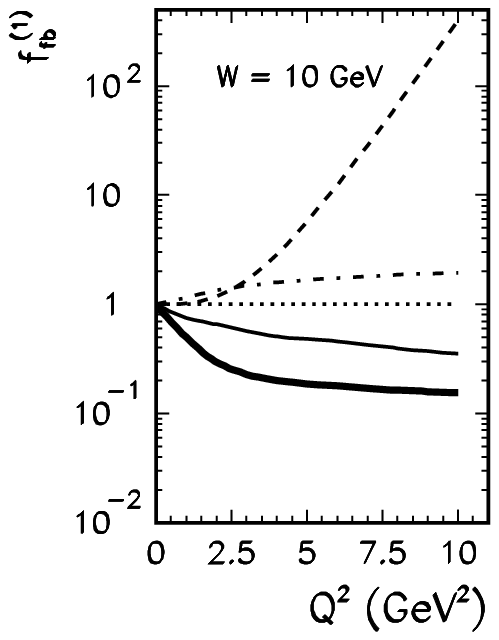}}
 \subfigure[]{\label{f1_q2_w100}
    \includegraphics[width=7cm]{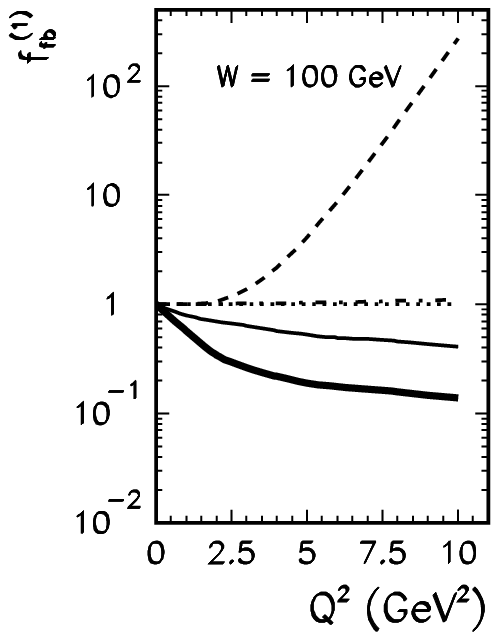}}
\caption{\it
Factorization breaking function $f_{fb}^{(1)}$ as a function 
of $Q^2$ ($Q^2 = Q_1^2 = Q_2^2$) for $W = 10$ GeV (left panel) 
and $W = 100$ GeV (right panel).
\label{fig:f1_q2_w}
}
\end{figure}

\begin{figure}[htb] % Figure 10
 \subfigure[]{\label{f2_q2_w10}
    \includegraphics[width=7cm]{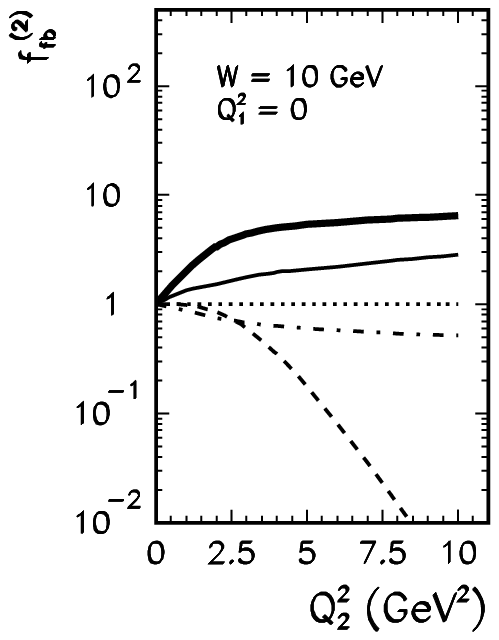}}
 \subfigure[]{\label{f2_q2_w100}
    \includegraphics[width=7cm]{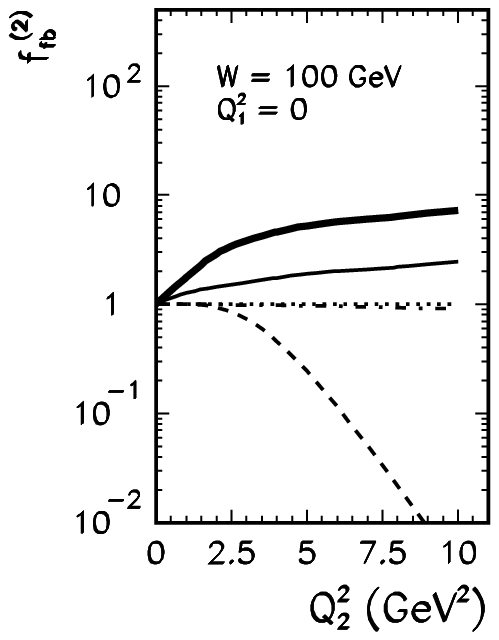}}
\caption{\it
Factorization breaking function $f_{fb}^{(2)}$ as a function 
of $Q_2^2$ ($Q_1^2 = 0$) for $W = 10$ GeV (left panel) 
and $W = 100$ GeV (right panel).
\label{fig:f2_q2_w}
}
\end{figure}

Having understood the general behaviour we wish to focus on
the most interesting parts of the $(Q_1^2,Q_2^2)$ space.
In Fig.\ref{fig:f1_q2_w} we show the behaviour of the two-dimensional
function $f_{fb}^{(1)}(Q_1^2,Q_2^2)$ along the diagonal
$Q^2 = Q_1^2 = Q_2^2$ and in Fig.\ref{fig:f2_q2_w} $f_{fb}^{(2)}(0,Q^2)
= f_{fb}^{(2)}(Q^2,0)$ along the line
$Q^2 = Q_2^2$  $ (Q_1^2 = 0)$. The thick solid line represents
our full model with all components included.
For illustration we have shown also factorization breaking functions
for separate mechanisms (components in the expansion (\ref{gg_expansion})).
Quite a different behaviour can be observed for different mechanisms.
Let us concentrate first on the $f_{fb}^{(1)}$ function.
While the single resolved and direct component grow with $Q^2$
the dipole-dipole component decreases. Paradoxically, the total 
$f_{fb}^{(1)}$ is smaller than the one for the dipole-dipole component.
This surprising result is related to the nonlinearity of
the quite complicated function $f_{fb}^{(1)}(Q^2,Q^2)$
which in fact involves four correlated points in the
$(Q_1^2,Q_2^2)$ plane.
A completely reverse behaviour can be seen for $f_{fb}^{(2)}$.
This has a simple analytic explanation. Substituting $Q_1^2 = Q^2$
and $Q_2^2 = Q^2$ into Eq.(\ref{f_fb1}) and $Q_1^2 = 0$ and $Q_2^2 = Q^2$
or $Q_1^2 = Q^2$ and $Q_2^2 = 0$ into Eq.(\ref{f_fb2}) we find:

%----------------------------------------------
\begin{equation}
f_{fb}^{(1)}(W,Q^2,Q^2) = \frac{1}{f_{fb}^{(2)}(W,0,Q^2)} =
\frac{1}{f_{fb}^{(2)}(W,Q^2,0)} \; .
\end{equation}
%----------------------------------------------

%--------------------
\section{Conclusions}
%--------------------

In our former paper we have constructed a simple model for
$\gamma^* p$ total cross section which, in contrast to other models in
the literature, includes the resolved photon component.
The latter is known to be the necessary ingredient when discussing exclusive
reactions.
In the present paper we have generalized the model to the case
of $\gamma \gamma$ scattering. In the last case a few new components appear so far not
discussed in the literature.

The naive generalization of our former model for $\gamma^* p$ total cross
section leads to a serious overestimation of the $\gamma \gamma$ total
cross sections. In general, this fact can be due either to a nonoptimal
set of model parameters found in our previous study or/and
due to some model simplifications. For instance, it is customary
that model parameters for resolved photon component
obtained in the vector meson dominance approach are taken from 
vector meson dileptonic decays, i.e. on 
meson mass shell. In the $\gamma^* p$ and $\gamma \gamma$ processes, 
of interest to us, vector mesons are rather off-shell. 
Therefore one could expect some off-shell effects. Calculating such 
off-shell effects in nonelementary processes is not a simple task. 
In this paper we have suggested to include such an effect 
by introducing new form factors which we call off-shell form factors 
for simplicity. On the other hand, when including the quark-antiquark 
continuum one usually takes into account the perturbative quark-antiquark 
"photon wave function". This is justified and reasonable for small
size dipoles only. 
The physics of large-size dipoles must involve nonperturbative 
effects which may lead to double counting in our model. In order 
to avoid double counting the large-size dipoles must be eliminated. 
We reduce their contribution using a simple exponential function 
in transverse dipole size. Summarizing, the two new functions 
bring in two new model parameters. Having this freedom we have performed 
a new fit of our generalized-model parameters to the $\gamma^*$p and 
$\gamma \gamma$ 
experimental data. The generalization of the model for meson off-shell
effects and large dipole size effects discussed above permits a 
simultaneous description of both processes considered.

When trying to extrapolate the experimental cross sections for
the $\gamma^* \gamma^*$ scattering to real photons one often 
assumes factorization.
Our multicomponent model violates this assumption.
We have quantified the effects of factorization breaking in our 
model with parameters fixed to describe the $\gamma^* p$ and 
$\gamma \gamma$ data. 
We have proposed two functions which can
be used as a measure of factorization breaking.
We have found a strong effect, rather weakly dependent of the
center of mass $\gamma^* \gamma^*$ energy. Experimental search
of such effects could teach us more about reaction
mechanism. Certainly, it is not an easy task with the LEP2 apparatus.

\vskip 1cm
%--------------------
{\bf Acknowledgments}
%--------------------
We are indebted to Mariusz Przybycie\'n from the OPAL collaboration
for an interesting discussion.

%---------------------------------------------------------

\newpage
%-------------------------------------------
%%%%%%%%%%%%%%%%%%%%%%%%%%%%%%%%%%%%%%%%%%%%

%-------------------------------
%--------------------------------
%--------------------------------
%-------------------------------
%-------------------------------
%------------------------------------
%------------------------------------
%------------------------------------
%------------------------------------
%----------------------------------
%----------------------------------
%----------------------------
%----------------------------
%----------------------------
%----------------------------
%--------------------------------
%--------------------------------
%--------------------------------
%--------------------------------
%--------------------------------

%--------------------------------
\end{document}